\documentclass[12pt, preprint]{emulateapj}
\usepackage{natbib}
\usepackage{enumerate}

\newcommand{\mass}{\mathcal{M}}

\newcommand{\refp}[1]{(\ref{#1})}
\newcommand{\Kepler}{\textit{Kepler}}

\newcommand{\icarus}{Icarus}

\newcommand{\figscale}{1}

\begin{document}
\shorttitle{II. Turbulence inhibits planetesimal accretion}

\shortauthors{Meschiari}

\title{Planet Formation in Circumbinary Configurations: Turbulence Inhibits Planetesimal Accretion}

\author{Stefano Meschiari\altaffilmark{1}}
\altaffiltext{1}{
McDonald Observatory, University of Texas at Austin, Austin, TX 78712
}
\email{stefano@astro.as.utexas.edu}

\begin{abstract}
The existence of planets born in environments highly perturbed by a stellar companion represents a major challenge to the paradigm of planet formation. In numerical simulations, the presence of a close binary companion stirs up the relative velocity between planetesimals, which is fundamental in determining the balance between accretion and erosion. However, the recent discovery of circumbinary planets by \Kepler{} establishes that planet formation in binary systems is clearly viable. We perform N-body simulations of planetesimals embedded in a protoplanetary disk, where planetesimal phasing is frustrated by the presence of stochastic torques, modeling the expected perturbations of turbulence driven by the magnetorotational instability (MRI). We examine perturbation amplitudes relevant to dead zones in the midplane (conducive to planet formation in single stars), and find that planetesimal accretion can be inhibited even in the outer disk (4-10 AU) far from the central binary, a location previously thought to be a plausible starting point for the formation of circumbinary planets.
\end{abstract}
\keywords{Planets and satellites: formation, Planets and satellites: dynamical evolution and stability}

\section{Introduction}
Among the numerous discoveries of the \Kepler{} mission, the detection of circumbinary (CB) planets Kepler 16 through the multiple system Kepler 47 \citep{Doyle11, Welsh12, Orosz12,  Orosz12a, Schwamb12} has  propelled a renewed theoretical effort in explaining how such planets would be assembled in a binary environment. In presence of a close binary companion, protoplanetary disks can become a rather hostile planetary nursery in both circumstellar and circumbinary configurations. The difficulties encountered by the standard core accretion paradigm in the binary environment are several, including truncation, mass loss and relatively fast dispersal of disks in close binaries \citep[e.g.][]{Duchene10, Kraus12}, possible vaporization of grains in dynamically excited disks \citep{Nelson00} and impaired planetesimal growth into protoplanets \citep{Marzari00, Moriwaki04, Thebault04, Thebault06, Scholl07, Paardekooper08, Thebault11}.

The latter ``planetesimal bottleneck'' is a robust consequence of the interplay between the gravitational perturbations of the stellar companion (which stirs the planetesimal disk and acts to raise eccentricities) and the aerodynamic drag from a putative protoplanetary disk. Since aerodynamic drag tends to both damp planetesimal eccentricities and align planetesimal orbits in a size-dependent fashion, planetesimals of different sizes will collide on different phases, resulting in high collisional speeds which lead to destructive (rather than accreting) events.
Recently, \citet{Meschiari12} (hereafter M12) investigated planetesimal accretion in the Kepler-16 system, using $N$-body simulations (coupled with drag from a static background disk) which track planetesimal collisions throughout a range of semi-major axes over $10^5$ years. The census of planetesimal collisions indicated that regions inside 4 AU ($\approx 20 a_B$, where $a_B$ is the semi-major axis of the central binary) were dominated by destructive events, and therefore hostile to planet formation. Consequently, we posited that Kepler 16 could have plausibly assembled outside the forbidden region and subsequently migrated inwards through tidal interaction with the protoplanetary disk, later stopping close to the inner edge of the disk. This scenario is supported by the evolution of planetary cores in hydrodynamical simulations \citep{Pierens07, Pierens08}, which also suggested Jupiter-mass planets in CB configurations should be rare (in accordance with the observed \Kepler{} sample). \citet{Paardekooper12} investigated planet formation in the Kepler 16, 34 and 35 systems using a similar approach, additionally including self-consistent planetesimal formation and destruction. They reached analogous conclusions, and asserted that \emph{in-situ} formation was unlikely, even under the most favorable conditions. 

The approach of M12, similarly to previous investigations, neglected several physical ingredients for the sake of computational expediency. Indeed, fully self-consistent simulations which include the hydrodynamical response of the protoplanetary disk have shown that additional oscillations in eccentricity and longitude of pericenter of the planetesimals might be introduced from the development of bulk eccentricity and spiral perturbations in the disk. However, their magnitude might depend somewhat on the details of the hydrodynamical simulation \citep[e.g.][]{Paardekooper08, Marzari12, Muller12}.

In this Letter, we consider magnetohydrodynamical turbulence driven by the magnetorotational instability (MRI) as an additional source of perturbations on the planetesimal disk. MRI-driven turbulence \citep{Balbus91} is thought to be the likely source of anomalous viscosity in protoplanetary disks \citep[e.g.][]{Armitage98}, and influence how planetesimal formation \citep[e.g.][]{Johansen07}, planetesimal accretion \citep[e.g.][]{Ogihara07, Ida08, Nelson10}, and planetary migration \citep[e.g.][]{Nelson03, Laughlin04, Nelson04, Baruteau10} proceed. Previous studies simulating the dynamics of planetesimals embedded in turbulent disks in single-star environments showed that in fully MRI-active disks, the velocity dispersion of the planetesimals is significantly raised by gravitational perturbations induced by density fluctuations. For typical disk parameters and nominal turbulence strength, km-sized planetesimals might be in a highly erosive regime. Such vigorous turbulence, however, might not be appropriate to the midplane of realistic protoplanetary disks, which is thought to be dominated by a ``dead zone'' with near-laminar flow \citep{Gammie96}. \citet{Gressel11} (hereafter G11) presented the results of  stratified, 3D MHD simulations (including a substantial dead zone); planetesimals embedded in the disk midplane at  5 AU experienced a significantly reduced excitation of their eccentricities (by a factor $\approx$ 10-20). Therefore, they concluded dead zone represent ``safe havens'' for the growth of km-sized planetesimals.

Although the reduced amplitude of the random velocities excited by turbulent fluctuations is potentially conducive to planetesimal accretion in the single-star environment, the situation is more complicated in the binary environment.  In the latter case, a very precise alignment between the planetesimal orbits is crucial to attaining low encounter speeds despite the substantial eccentricity of the planetesimal orbits. Random kicks diffuse planetesimals out of alignment; although aerodynamic drag would attempt to restore alignment, it will do so on a timescale that is size-dependent, once again differentially phasing planetesimals of different sizes. Finally, high-frequency radial oscillations of the planetesimal eccentricity (as would be caused by stochastic torques) would potentially lead to further orbital crossing, leading to high encounter velocities. Therefore, we anticipate that planet formation might be strongly perturbed, or even inhibited, despite the smaller turbulent amplitudes appropriate to the midplane dead zone. We couple the $N$-body code of M12 with a numerical model that approximates the stochastic torquing arising from MRI turbulence. Our analysis shows that this additional source of perturbations is potentially damaging to planet formation in the outer disk, which was previously thought to be relatively protected from destructive impacts and therefore a plausible location for core assembly \citep[M12;][]{Paardekooper12}.

The plan of this Letter is as follows. In \S \ref{sec:setup}, we briefly discuss our numerical model for the gravitational torques arising from MHD turbulence. In \S \ref{sec:results} we show the results of our simulations, and discuss them in the context of planet formation in \S \ref{sec:conc}.

\section{Numerical setup}\label{sec:setup}
In this Letter, we study planet formation by sampling planetesimal collision events between 4 and 10 AU, and compare them to velocity thresholds corresponding to destructive impacts. We consider the orbital elements of the Kepler-16 system as our prototypical configuration, and to facilitate comparison with previous investigations \citep[M12;][]{Paardekooper12}. We refer the reader to M12 for a description of the numerical code, initial setup and velocity thresholds. 

\subsection{Turbulent model}
\begin{figure}
\epsscale{\figscale}
\plotone{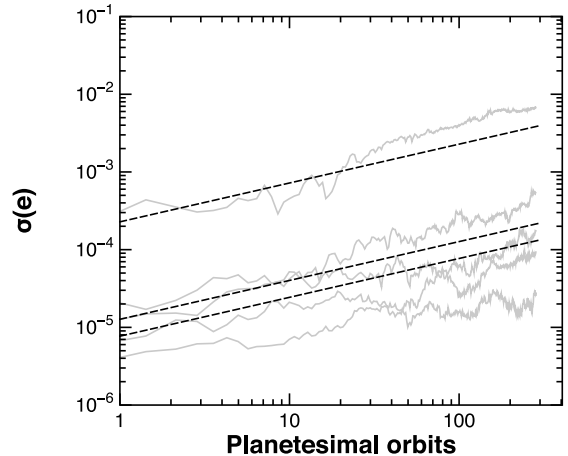}
\caption{Sample growth plot of the eccentricity dispersion for models A, B1-4 (grey lines, from top to bottom). The dashed lines indicate the best-fit eccentricity dispersion measured in the full MHD simulations of G11, models A, D1 and D2 (from top to bottom).}\label{fig:turb_ecc}
\end{figure}
We use the analytical prescription of \citet{Laughlin04} (hereafter L04) to model torques arising from density fluctuations in the disk; we include corrections to the formulation from \citet{Ogihara07} \citep[also used in ][]{Ida08, Baruteau10}, and introduce small modifications that reduce its computational cost. In this model, density fluctuations are forced by a potential $\Phi$, a sum of turbulent $m$-fold modes
\begin{equation}
\Phi = \gamma r^2 \Omega^2 \sum_i \xi_i R(r, r_{c, i}, \sigma_i)\ T(\tilde{t}_i)\ \cos(m\varphi - \varphi_{c, i} - \Omega_{c, i} \tilde{t}_i)\label{eqn:turb}
\end{equation}
where $\gamma$ sets the overall turbulent amplitude, $r_{c, i}$, $\varphi_{c, i}$ and $\Omega_{c, i}$ are the radial center, phase and angular velocity of the mode (picked randomly in the disk), $m$ is sampled from a lograndom distribution between 1 and 6, $\sigma_i = \pi r_{c, i} / 4m$ is the radial extent of the mode, $\xi_i$ is sampled from a Gaussian distribution of unit variance, $\tilde{t}_i$ is the lifetime of the mode normalized by a timescale $\Delta t_i$ and $R$ and $T$ are two Gaussian-like functions centered around $r_{c, i}$ and $0.5 \Delta t_i$, respectively. Each mode has a limited lifetime $\Delta t_i$; following \citet{Baruteau10}, we reduce the lifetimes by a factor of 10 from the prescription of L04 in order to better match the autocorrelation timescale of 3D MHD simulations.  

The actual force on the planetesimals will arise from the gravitational force of the density fluctuations induced by Equation \refp{eqn:turb}.  In order to proceed without the full hydrodynamical machinery, L04 used a WKB analysis to derive the following scaling for the RMS torque on a planetesimal of mass $\mass_{pl}$:
\begin{equation}
\tau_T = C \gamma r^4 \Omega^2\Sigma_g \frac{\mass_{pl}}{\mass_*}\ \label{eqn:torque}
\end{equation}
where $\mass_*$ is the mass of the central object (in our case, the total binary mass) and $\Sigma_g$ is the unperturbed surface density of the protoplanetary disk. While L04 also derives an approximate value for the constant $C$, the heuristic nature of the derivation \citep[which is equivalent to a dimensional analysis, as derived by ][]{Johnson06} suggests a better approach would be to fit the amplitude of the perturbations $\gamma$ to the results of MHD simulations.

To compare with the results of G11, we integrated the orbits of an ensemble of 100 planetesimals, started at 5 AUs with zero eccentricity and random initial phase, subject to the stochastic torques described above and no aerodynamic drag and computed the diffusion of their orbital elements. We consider several models, differing only by the turbulent amplitude $\gamma$: model A ($\gamma = 2.5\times 10^{-3}$) represents a fiducial turbulent amplitude appropriate to fully MRI-active disks, while models B1-4 ($\gamma = 2.5\times 10^{-4}$, $10^{-4}$, $5 \times 10^{-5}$, $2.5 \times 10^{-5}$, respectively) have reduced turbulent amplitudes appropriate to the midplane of dead zones.  Since $\alpha$ scales as $\gamma^2$ \citep[where $\alpha$ is the usual viscosity parameter in the Shakura-Sunyaev prescription; ][]{Baruteau10}, the values chosen span two orders of magnitude of turbulent viscosity. Finally, for comparison with M12, we also ran simulations with no turbulence (model C). 

 Figure \ref{fig:turb_ecc} shows the growth of the dispersion of the planetesimal eccentricity $\sigma(e)$ as a function of time; as expected from a random-walk process, $\sigma(e) \propto t^{1/2}$. Models B1 and B2 bracket well the growth of the eccentricity dispersion seen by G11 for models with dead zones; we include the reduced values of models B3 and B4 to test the robustness of our results. 

We remark that it is likely that the interaction between the planetesimals and the background disk is more complicated than the model presented in this Letter; for instance, we expect the disk to be endowed with some eccentricity \citep[e.g.][]{Marzari08, Marzari12}. However, for the sake of simplicity and to highlight the role of turbulence as an additional factor in dephasing planetesimal orbits, we decided to take the orthogonal approach of ignoring the self-consistent evolution of the disk. In this picture, we assume a circular disk as a maximally accretion-friendly starting point.  

\section{Simulations}\label{sec:results}
\subsection{Diffusion of eccentricity and dephasing}

\begin{figure}
\epsscale{\figscale}
\plotone{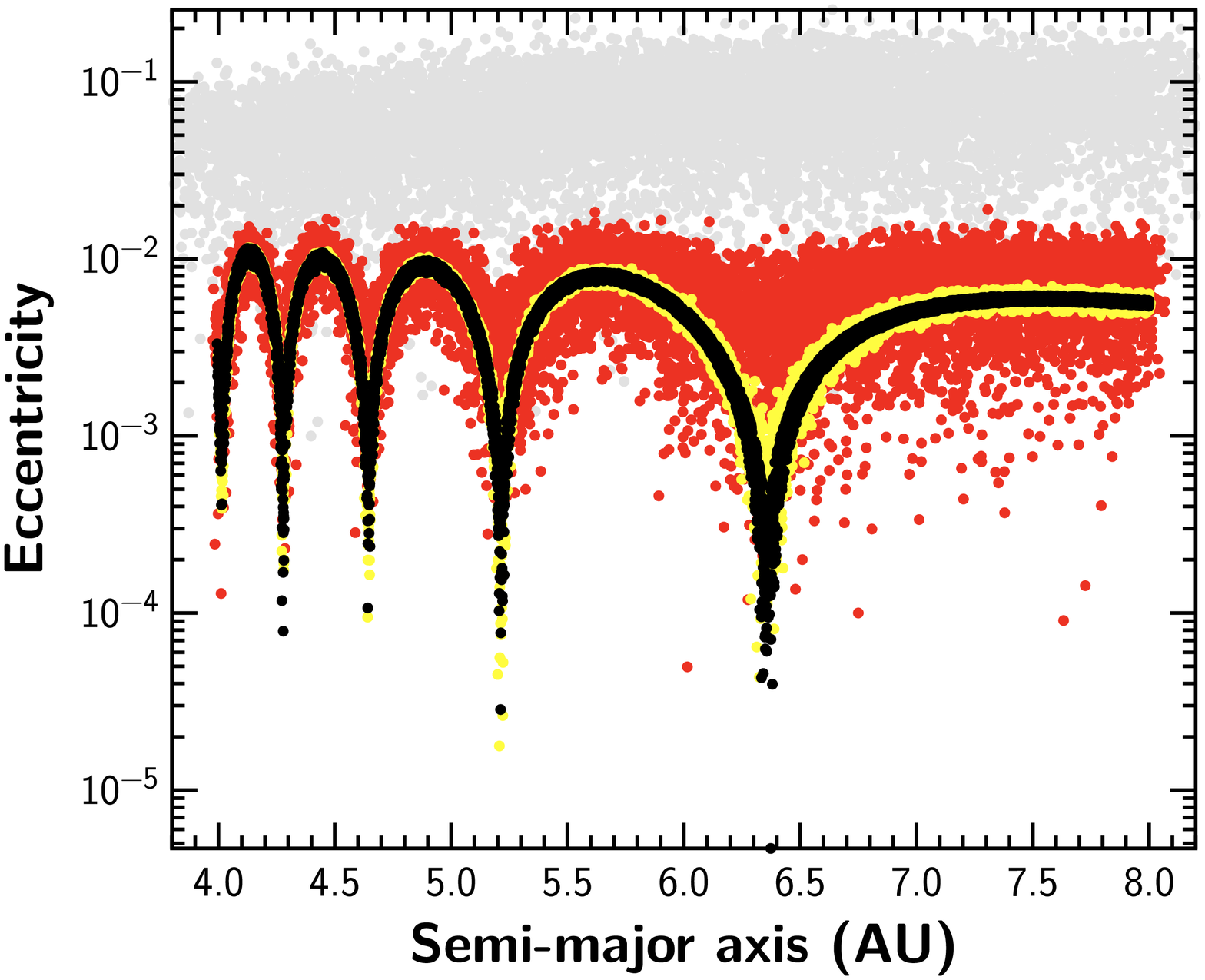}
\plotone{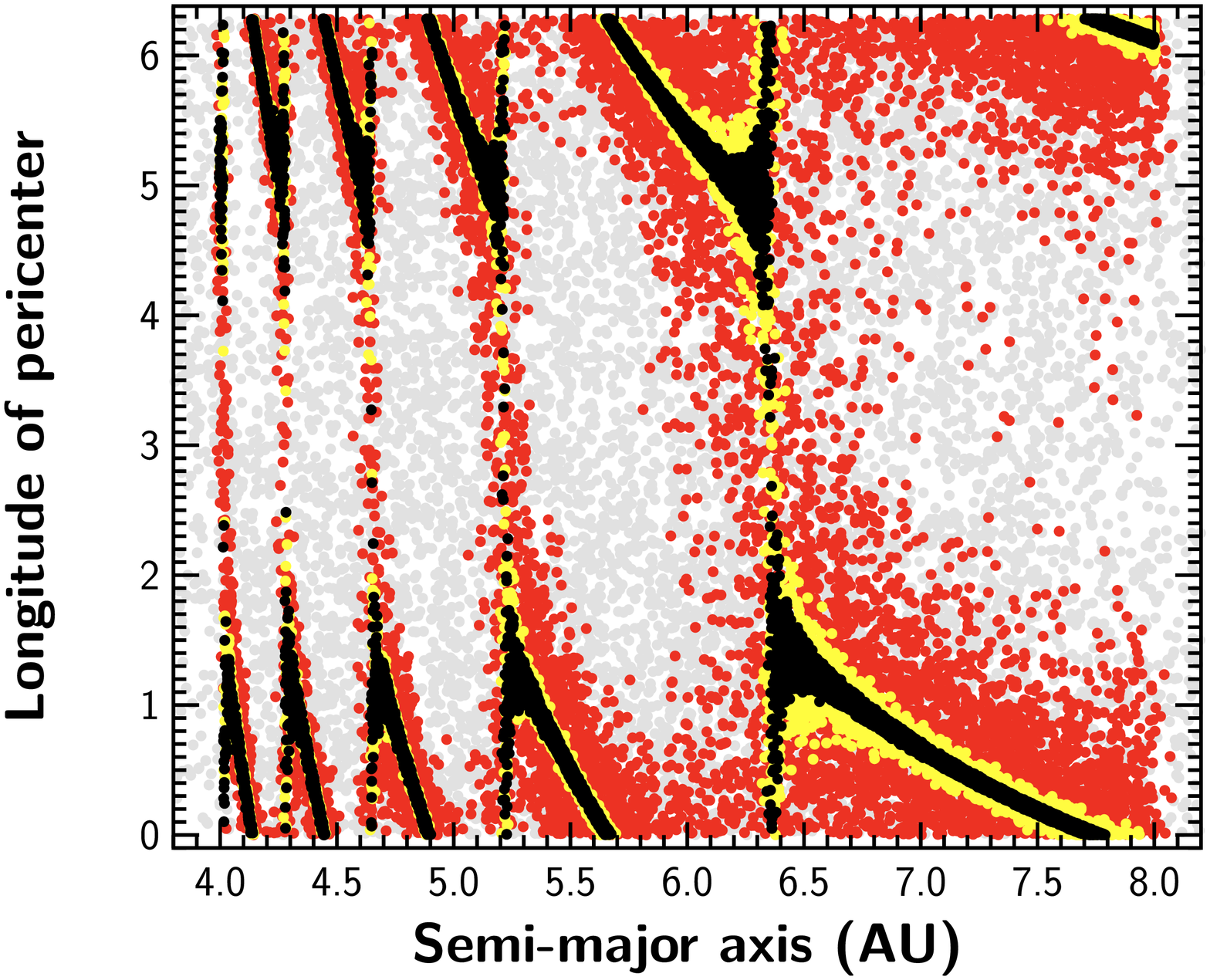}
\caption{Eccentricity and longitude of pericenter $\varpi - \varpi_B$ as a function of semi-major axis for 10,000 planetesimals distributed between 4 and 8 AU, for runs with no turbulence (black dots), model A (grey dots, active-MRI disk), model B1 (red dots, dead zone) and model B4 (yellow dots, dead zone turbulence amplitude reduced by a factor of 10). }\label{fig:ecclop}
\end{figure}

To visualize the effect of the turbulent fluctuation on a population of planetesimal, we first integrated the trajectories of a swarm of 10,000 planetesimals uniformly distributed between 4 and 8 AUs for four different levels of turbulence: no turbulence, model A (active MRI-disk), model B1 (nominal dead zone turbulence fitting the results of G11) and model B4 (turbulence amplitude reduced by a factor of 10). 

Figure \ref{fig:ecclop} shows the eccentricity and longitude of pericenter ($\varpi$) of the planetesimals. Compared to the run without turbulence, both models with levels of turbulence appropriate to a dead zone add a considerable amount of noise on top of the $e$ and $\varpi$-profiles secularly imposed by the central binary. Since the orbits of neighboring planetesimals will not be as collimated, impact velocities will be raised everywhere in the disk. Therefore, we expect that the fraction of accreting impacts will be decreased even when very low levels of turbulence are present. 

For the model representing a fully MRI active disk, eccentricities are raised to very high values ($e \approx 5\times 10^{-2}$) and the longitude of pericenter is completely randomized; therefore, the collision velocity can be directly estimated as $\Delta v \approx e v_{kep} > 400$ m/s, comfortably above any velocity threshold resulting in destructive impacts \citep{Stewart09}. Indeed, in this case the turbulent torques completely overwhelm the secular forcing of the central binary, resulting in high impact speeds that are consistent with the results of simulations of planetesimal dynamics in fully turbulent disks around single stars \citep[e.g.][]{Ida08, Nelson10}.

\subsection{Collision statistics}

We subsequently ran full simulations with collision detection between 4 and 10 AU, for each of our models. Given the computational overhead of calculating and updating the turbulent forcing at each time step, we chose to instead integrate the trajectories of a smaller number of planetesimals (1,000) concentrated in annuli centered around 4, 6, 8 and 10 AUs. This approach has the advantage of reducing the running time of our simulations, while simultaneously improving our collision statistics by increasing the impact rate.

\begin{figure}
\epsscale{\figscale}
\plotone{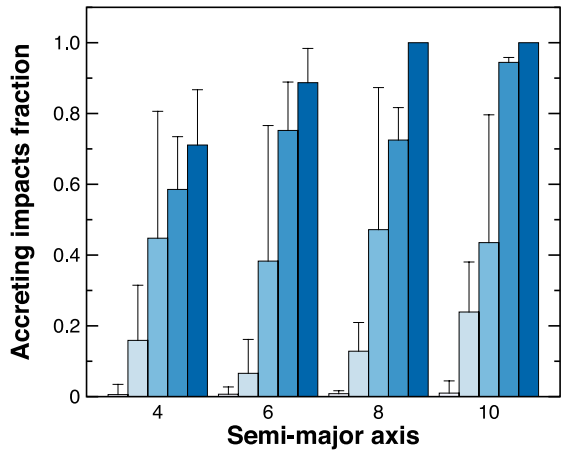}
\caption{Fraction of accreting impacts as a function of semi-major axis (from lighter to darker: models B1-B4 and C.) Each bar represents the percentage of unperturbed impacts; the fraction of perturbed impacts is additionally represented by a solid line.}\label{fig:frac}
\end{figure}

Figure \ref{fig:frac} shows the fraction of accreting impacts as a function of semi-major axis for each of our models. For the sake of completeness, we further classify impacts as ``unperturbed'' ($\Delta v < v_{esc}$, where $v_{esc}$ is the escape velocity) or ``perturbed'' accretion ($v_{esc} < \Delta v < v_{ero}$); for the former, the impact velocity is low enough to allow runaway growth to proceed, therefore allowing the rapid formation of oligarchs. We generously deem a radial bin as a favorable location for planet formation when the fraction of accreting impacts (unperturbed or perturbed) is larger than 50\%.

For models B1 and B2 (with turbulent amplitudes which best fit the results of the 3D MHD simulations of G11), we find that the percentage of accreting impacts decreases dramatically. This is consistent with the noisiness of the pericenter phasing and the high-frequency jitter around the damped value. These short-term oscillations are not damped efficiently by the aerodynamic drag at these large distances, since the drag torque decreases steeply with distance ($\tau_G \propto a^{-3-\beta}$, for planetesimals at the equilibrium forced eccentricity) while the turbulent torque in our model decreases more slowly ($\tau_{T} \propto a^{1-\beta}$), where we take $\Sigma \propto r^{-\beta}$ ($\beta = 1.75$ for the minimum-mass solar nebula model considered in this paper). 

For models B3 and B4 (with reduced turbulent amplitudes) we find that, despite the fact that substantial oscillations in eccentricity and longitude of pericenter are still induced, the resulting impact velocities have crossed the accreting threshold. The median impact speed for model B4 is $\approx 6$ m/s, which is in the erosive regime for 1-km planetesimals but allow accretion for 10-km planetesimals.

Finally, in accordance with the simulations of M12, we find that planet formation can proceed undisturbed outside 4 AU when turbulent forcing is switched off. 

\section{Discussion}\label{sec:conc}
In this Letter we have investigated planetesimal accretion in the outer parts of a circumbinary disk (4-10 AU). Our simulations indicate that the stochastic forcing of turbulent perturbations will frustrate planetesimal phasing and raise eccentricities, inhibiting planet formation even quite far from the central binary. This result is robust for levels of turbulence that match those observed in realistic MHD simulations. In runs modeling disks fully invaded by MRI turbulence, planetesimal phasing is completely destroyed and planetesimals will collide at speeds that are much higher than fiducial erosive velocities. We then ran simulations fitting the turbulent amplitude corresponding to the stratified disks endowed with a large dead zone obtained in G11, resulting in a reduction by a factor $\approx 10-20$ in the turbulent torque. While the reduced stochastic perturbations would be sufficiently small to allow accretion in a single-star environment, in our circumbinary configuration the fraction of impacts resulting in accreting events was greatly reduced ($\lesssim 1-30\%$ between 4 and 10 AU for models B1 and B2, respectively) due to the increased eccentricity jitter and dephasing. Therefore, it is possible to inhibit planet formation by two different mechanisms: by high eccentricities, differential phasing, and perturbations caused by the self-consistent reaction of the gas disk at small radii, and by stochastic turbulent torques at large radii. This is problematic, since if that picture is accurate, it is not feasible to form the observed planetary census at any realistic distance from the binary. The circumbinary environment is less robust than single stars with regards to dynamical perturbations induced by the disk (regardless of their origins); a further reduced level of turbulence may be required to form planets at all. Indeed, we found that further reduction of the turbulent amplitude by a factor of 2-5 was necessary to make planet formation viable again in the outer disk. 

Several uncertainties in our model may still provide some room for allowing planet formation despite the perturbing effects of turbulence. Our analytic prescription, while attempting to match MHD simulations, only provides a ``0-th'' order description of gravitational perturbations induced by torques. For instance, we remark that we tuned the amplitude of turbulent torques based on the evolution of swarm of planetesimals in a single MHD simulation, and rescaled the turbulent amplitude at each radial location according to the numerical prescription of Section \ref{sec:setup}; however, the output of these models will depend on the assumed magnetic field, the ionization level and possibly the resolution of the shearing box. 
The detailed radial dependence and even extent of the dead zone remain uncertain and depend on the assumed model \citep[e.g.][; if the dead zone extends to only a few AUs, then fully active turbulence may still play a role in the outer disk and even shut off planetesimal accretion completely]{Matsumura05, Terquem08, Flaig12}. Effectively, using the torque scaling of Equation \ref{eqn:torque} implicitly assumes that the turbulent torque is well described by the fully turbulent model of L04, attenuated by a constant factor at each radial location. Only self-consistent (including the time-dependent potential of the binary), computationally expensive global simulations could, in principle, inform the model presented in this Letter. However, since we have considered various values of $\gamma$ for each independent annulus, we are reasonably covering a number of possibilities for the radial dependence of the turbulence amplitude. Our results at a given radius should only be considered suggestive of further challenges to planet formation.

Finally, we remark that, similarly to M12, we have chosen to study planetesimal growth through mutual collisions in the 1-10 km planetesimal regime. A more sophisticated approach following the evolution of the size distribution of the planetesimals (adding a significant computational overhead) may be warranted; in the latter approach, coevolution with a disk of small dust might make accretion possible closer to the binary \citep[assuming a high dust accretion efficiency; ][]{Paardekooper12}. Even in this setup, some amount of migration is needed to bring the fully formed core (or embryos) to the current location. The assumption of a static gas background is not realistic, especially in the inner disk; in that case, self-consistent gas dynamics will act to further increase impact velocities in the inner disk \citep{Paardekooper08}. We plan to incorporate the full hydrodynamical evolution of the gas disk in a follow-up paper using the SPH module of our code.

Our simulations indicate that planetesimal accretion in the 1-10 km range will be inhibited everywhere in the disk, and could only proceed if the strength of the turbulent torques is reduced from our fiducial value, or the initial planetesimal population is comprised of bigger objects. The latter scenario is particularly appealing, as it could allow planet formation to proceed in other highly dynamically disturbed environments \citep[e.g.][]{Thebault11}. Indeed, our fiducial model for dead zone turbulence becomes accretion-friendly for a planetesimal size spectrum spanning 10-100 km size. Recent simulations of planetesimal formation in weakly turbulent disks show that massive bound clumps are formed rapidly from meter-sized boulders within pressure bumps, for typical nebula parameters \citep[e.g. ][]{Johansen07, Johansen11}. Such bound clumps will likely result in planetesimals with sizes comparable to at least a substantial fraction of the dwarf planet Ceres. We suggest that a primordial population of large planetesimals might be crucial to proceed with planet formation in highly perturbed environments. However, further simulations will be required to assess whether the formation of large clumps is robust to the dynamical perturbations of a binary companion.

\acknowledgments
S.M. acknowledges insightful discussions with G. Laughlin and N. Haghighipour, a useful critique from the anonymous referee, as well as support from the W. J. McDonald Postdoctoral Fellowship.

\bibliographystyle{apj}

\begin{thebibliography}{38}
\bibitem[{{Armitage}(1998)}]{Armitage98}
{Armitage}, P.~J. 1998, \apjl, 501, L189

\bibitem[{{Balbus} \& {Hawley}(1991)}]{Balbus91}
{Balbus}, S.~A., \& {Hawley}, J.~F. 1991, \apj, 376, 214

\bibitem[{{Baruteau} \& {Lin}(2010)}]{Baruteau10}
{Baruteau}, C., \& {Lin}, D.~N.~C. 2010, \apj, 709, 759

\bibitem[{{Cuzzi} {et~al.}(2008){Cuzzi}, {Hogan}, \& {Shariff}}]{Cuzzi08}
{Cuzzi}, J.~N., {Hogan}, R.~C., \& {Shariff}, K. 2008, \apj, 687, 1432

\bibitem[{{Doyle} {et~al.}(2011){Doyle}, {Carter}, {Fabrycky}, {Slawson},
  {Howell}, {Winn}, {Orosz}, {Prsa}, {Welsh}, {Quinn}, {Latham}, {Torres},
  {Buchhave}, {Marcy}, {Fortney}, {Shporer}, {Ford}, {Lissauer}, {Ragozzine},
  {Rucker}, {Batalha}, {Jenkins}, {Borucki}, {Koch}, {Middour}, {Hall},
  {McCauliff}, {Fanelli}, {Quintana}, {Holman}, {Caldwell}, {Still},
  {Stefanik}, {Brown}, {Esquerdo}, {Tang}, {Furesz}, {Geary}, {Berlind},
  {Calkins}, {Short}, {Steffen}, {Sasselov}, {Dunham}, {Cochran}, {Boss},
  {Haas}, {Buzasi}, \& {Fischer}}]{Doyle11}
{Doyle}, L.~R., {Carter}, J.~A., {Fabrycky}, D.~C., {Slawson}, R.~W., {Howell},
  S.~B., {Winn}, J.~N., {Orosz}, J.~A., {Prsa}, A., {Welsh}, W.~F., {Quinn},
  S.~N., {Latham}, D., {Torres}, G., {Buchhave}, L.~A., {Marcy}, G.~W.,
  {Fortney}, J.~J., {Shporer}, A., {Ford}, E.~B., {Lissauer}, J.~J.,
  {Ragozzine}, D., {Rucker}, M., {Batalha}, N., {Jenkins}, J.~M., {Borucki},
  W.~J., {Koch}, D., {Middour}, C.~K., {Hall}, J.~R., {McCauliff}, S.,
  {Fanelli}, M.~N., {Quintana}, E.~V., {Holman}, M.~J., {Caldwell}, D.~A.,
  {Still}, M., {Stefanik}, R.~P., {Brown}, W.~R., {Esquerdo}, G.~A., {Tang},
  S., {Furesz}, G., {Geary}, J.~C., {Berlind}, P., {Calkins}, M.~L., {Short},
  D.~R., {Steffen}, J.~H., {Sasselov}, D., {Dunham}, E.~W., {Cochran}, W.~D.,
  {Boss}, A., {Haas}, M.~R., {Buzasi}, D., \& {Fischer}, D. 2011, Science, 333,
  1602

\bibitem[{{Duch{\^e}ne}(2010)}]{Duchene10}
{Duch{\^e}ne}, G. 2010, \apjl, 709, L114

\bibitem[Flaig et al.(2012)]{Flaig12} Flaig, M., Ruoff, P., 
Kley, W., \& Kissmann, R.\ 2012, \mnras, 420, 2419 

\bibitem[{{Gammie}(1996)}]{Gammie96}
{Gammie}, C.~F. 1996, \apj, 457, 355

\bibitem[{{Gressel} {et~al.}(2011){Gressel}, {Nelson}, \& {Turner}}]{Gressel11}
{Gressel}, O., {Nelson}, R.~P., \& {Turner}, N.~J. 2011, \mnras, 415, 3291

\bibitem[{{Ida} {et~al.}(2008){Ida}, {Guillot}, \& {Morbidelli}}]{Ida08}
{Ida}, S., {Guillot}, T., \& {Morbidelli}, A. 2008, \apj, 686, 1292

\bibitem[{{Johansen} {et~al.}(2011){Johansen}, {Klahr}, \&
  {Henning}}]{Johansen11}
{Johansen}, A., {Klahr}, H., \& {Henning}, T. 2011, \aap, 529, A62

\bibitem[{{Johansen} {et~al.}(2007){Johansen}, {Oishi}, {Mac Low}, {Klahr},
  {Henning}, \& {Youdin}}]{Johansen07}
{Johansen}, A., {Oishi}, J.~S., {Mac Low}, M.-M., {Klahr}, H., {Henning}, T.,
  \& {Youdin}, A. 2007, \nat, 448, 1022

\bibitem[{{Johnson} {et~al.}(2006){Johnson}, {Goodman}, \& {Menou}}]{Johnson06}
{Johnson}, E.~T., {Goodman}, J., \& {Menou}, K. 2006, \apj, 647, 1413

\bibitem[{{Kraus} {et~al.}(2012){Kraus}, {Ireland}, {Hillenbrand}, \&
  {Martinache}}]{Kraus12}
{Kraus}, A.~L., {Ireland}, M.~J., {Hillenbrand}, L.~A., \& {Martinache}, F.
  2012, \apj, 745, 19

\bibitem[{{Laughlin} {et~al.}(2004){Laughlin}, {Steinacker}, \&
  {Adams}}]{Laughlin04}
{Laughlin}, G., {Steinacker}, A., \& {Adams}, F.~C. 2004, \apj, 608, 489

\bibitem[Matsumura 
\& Pudritz(2005)]{Matsumura05} Matsumura, S., \& Pudritz, R.~E.\ 2005, \apjl, 618, L137 

\bibitem[Marzari et al.(2008)]{Marzari08} Marzari, F., 
Th{\'e}bault, P., \& Scholl, H.\ 2008, \apj, 681, 1599 

\bibitem[{{Marzari} {et~al.}(2012){Marzari}, {Baruteau}, {Scholl}, \&
  {Thebault}}]{Marzari12}
{Marzari}, F., {Baruteau}, C., {Scholl}, H., \& {Thebault}, P. 2012, \aap, 539,
  A98

\bibitem[{{Marzari} \& {Scholl}(2000)}]{Marzari00}
{Marzari}, F., \& {Scholl}, H. 2000, \apj, 543, 328

\bibitem[{{Meschiari}(2012)}]{Meschiari12}
{Meschiari}, S. 2012, \apj, 752, 71

\bibitem[{{Moriwaki} \& {Nakagawa}(2004)}]{Moriwaki04}
{Moriwaki}, K., \& {Nakagawa}, Y. 2004, \apj, 609, 1065

\bibitem[{{M{\"u}ller} \& {Kley}(2012)}]{Muller12}
{M{\"u}ller}, T.~W.~A., \& {Kley}, W. 2012, \aap, 539, A18

\bibitem[{{Nelson}(2000)}]{Nelson00}
{Nelson}, A.~F. 2000, \apjl, 537, L65

\bibitem[{{Nelson} \& {Gressel}(2010)}]{Nelson10}
{Nelson}, R.~P., \& {Gressel}, O. 2010, \mnras, 409, 639

\bibitem[{{Nelson} \& {Papaloizou}(2003)}]{Nelson03}
{Nelson}, R.~P., \& {Papaloizou}, J.~C.~B. 2003, \mnras, 339, 993

\bibitem[{{Nelson} \& {Papaloizou}(2004)}]{Nelson04}
---. 2004, \mnras, 350, 849

\bibitem[{{Ogihara} {et~al.}(2007){Ogihara}, {Ida}, \&
  {Morbidelli}}]{Ogihara07}
{Ogihara}, M., {Ida}, S., \& {Morbidelli}, A. 2007, \icarus, 188, 522

\bibitem[{{Orosz} {et~al.}(2012{\natexlab{a}}){Orosz}, {Welsh}, {Carter},
  {Brugamyer}, {Buchhave}, {Cochran}, {Endl}, {Ford}, {MacQueen}, {Short},
  {Torres}, {Windmiller}, {Agol}, {Barclay}, {Caldwell}, {Clarke}, {Doyle},
  {Fabrycky}, {Geary}, {Haghighipour}, {Holman}, {Ibrahim}, {Jenkins},
  {Kinemuchi}, {Li}, {Lissauer}, {Prsa}, {Ragozzine}, {Shporer}, {Still}, \&
  {Wade}}]{Orosz12}
{Orosz}, J.~A., {Welsh}, W.~F., {Carter}, J.~A., {Brugamyer}, E., {Buchhave},
  L.~A., {Cochran}, W.~D., {Endl}, M., {Ford}, E.~B., {MacQueen}, P., {Short},
  D.~R., {Torres}, G., {Windmiller}, G., {Agol}, E., {Barclay}, T., {Caldwell},
  D.~A., {Clarke}, B.~D., {Doyle}, L.~R., {Fabrycky}, D.~C., {Geary}, J.~C.,
  {Haghighipour}, N., {Holman}, M.~J., {Ibrahim}, K.~A., {Jenkins}, J.~M.,
  {Kinemuchi}, K., {Li}, J., {Lissauer}, J.~J., {Prsa}, A., {Ragozzine}, D.,
  {Shporer}, A., {Still}, M., \& {Wade}, R.~A. 2012{\natexlab{a}}, ArXiv
  e-prints

\bibitem[{{Orosz} {et~al.}(2012{\natexlab{b}}){Orosz}, {Welsh}, {Carter},
  {Fabrycky}, {Cochran}, {Endl}, {Ford}, {Haghighipour}, {MacQueen}, {Mazeh},
  {Sanchis-Ojeda}, {Short}, {Torres}, {Agol}, {Buchhave}, {Doyle}, {Isaacson},
  {Lissauer}, {Marcy}, {Shporer}, {Windmiller}, {Barclay}, {Boss}, {Clarke},
  {Fortney}, {Geary}, {Holman}, {Huber}, {Jenkins}, {Kinemuchi}, {Kruse},
  {Ragozzine}, {Sasselov}, {Still}, {Tenenbaum}, {Uddin}, {Winn}, {Koch}, \&
  {Borucki}}]{Orosz12a}
{Orosz}, J.~A., {Welsh}, W.~F., {Carter}, J.~A., {Fabrycky}, D.~C., {Cochran},
  W.~D., {Endl}, M., {Ford}, E.~B., {Haghighipour}, N., {MacQueen}, P.~J.,
  {Mazeh}, T., {Sanchis-Ojeda}, R., {Short}, D.~R., {Torres}, G., {Agol}, E.,
  {Buchhave}, L.~A., {Doyle}, L.~R., {Isaacson}, H., {Lissauer}, J.~J.,
  {Marcy}, G.~W., {Shporer}, A., {Windmiller}, G., {Barclay}, T., {Boss},
  A.~P., {Clarke}, B.~D., {Fortney}, J., {Geary}, J.~C., {Holman}, M.~J.,
  {Huber}, D., {Jenkins}, J.~M., {Kinemuchi}, K., {Kruse}, E., {Ragozzine}, D.,
  {Sasselov}, D., {Still}, M., {Tenenbaum}, P., {Uddin}, K., {Winn}, J.~N.,
  {Koch}, D.~G., \& {Borucki}, W.~J. 2012{\natexlab{b}}, ArXiv e-prints

\bibitem[{{Paardekooper} {et~al.}(2012){Paardekooper}, {Leinhardt}, {Thebault},
  \& {Baruteau}}]{Paardekooper12}
{Paardekooper}, S.-J., {Leinhardt}, Z.~M., {Thebault}, P., \& {Baruteau}, C.
  2012, ArXiv e-prints

\bibitem[{Paardekooper {et~al.}(2008)Paardekooper, Th{\'e}bault, \&
  Mellema}]{Paardekooper08}
Paardekooper, S.-J., Th{\'e}bault, P., \& Mellema, G. 2008, Monthly Notices of
  the Royal Astronomical Society, 386, 973

\bibitem[{{Pierens} \& {Nelson}(2007)}]{Pierens07}
{Pierens}, A., \& {Nelson}, R.~P. 2007, \aap, 472, 993

\bibitem[{{Pierens} \& {Nelson}(2008)}]{Pierens08}
---. 2008, \aap, 483, 633

\bibitem[{{Scholl} {et~al.}(2007){Scholl}, {Marzari}, \&
  {Th{\'e}bault}}]{Scholl07}
{Scholl}, H., {Marzari}, F., \& {Th{\'e}bault}, P. 2007, \mnras, 380, 1119

\bibitem[Schwamb et al.(2012)]{Schwamb12} Schwamb, M.~E., Orosz, 
J.~A., Carter, J.~A., et al.\ 2012, arXiv:1210.3612 

\bibitem[{{Stewart} \& {Leinhardt}(2009)}]{Stewart09}
{Stewart}, S.~T., \& {Leinhardt}, Z.~M. 2009, \apjl, 691, L133

\bibitem[Terquem(2008)]{Terquem08} Terquem, C.~E.~J.~M.~L.~J.\ 
2008, \apj, 689, 532 

\bibitem[{{Thebault}(2011)}]{Thebault11}
{Thebault}, P. 2011, Celestial Mechanics and Dynamical Astronomy, 111, 29

\bibitem[{{Th{\'e}bault} {et~al.}(2006){Th{\'e}bault}, {Marzari}, \&
  {Scholl}}]{Thebault06}
{Th{\'e}bault}, P., {Marzari}, F., \& {Scholl}, H. 2006, \icarus, 183, 193

\bibitem[{{Th{\'e}bault} {et~al.}(2004){Th{\'e}bault}, {Marzari}, {Scholl},
  {Turrini}, \& {Barbieri}}]{Thebault04}
{Th{\'e}bault}, P., {Marzari}, F., {Scholl}, H., {Turrini}, D., \& {Barbieri},
  M. 2004, \aap, 427, 1097

\bibitem[{{Welsh} {et~al.}(2012){Welsh}, {Orosz}, {Carter}, {Fabrycky}, {Ford},
  {Lissauer}, {Pr{\v s}a}, {Quinn}, {Ragozzine}, {Short}, {Torres}, {Winn},
  {Doyle}, {Barclay}, {Batalha}, {Bloemen}, {Brugamyer}, {Buchhave},
  {Caldwell}, {Caldwell}, {Christiansen}, {Ciardi}, {Cochran}, {Endl},
  {Fortney}, {Gautier}, {Gilliland}, {Haas}, {Hall}, {Holman}, {Howard},
  {Howell}, {Isaacson}, {Jenkins}, {Klaus}, {Latham}, {Li}, {Marcy}, {Mazeh},
  {Quintana}, {Robertson}, {Shporer}, {Steffen}, {Windmiller}, {Koch}, \&
  {Borucki}}]{Welsh12}
{Welsh}, W.~F., {Orosz}, J.~A., {Carter}, J.~A., {Fabrycky}, D.~C., {Ford},
  E.~B., {Lissauer}, J.~J., {Pr{\v s}a}, A., {Quinn}, S.~N., {Ragozzine}, D.,
  {Short}, D.~R., {Torres}, G., {Winn}, J.~N., {Doyle}, L.~R., {Barclay}, T.,
  {Batalha}, N., {Bloemen}, S., {Brugamyer}, E., {Buchhave}, L.~A., {Caldwell},
  C., {Caldwell}, D.~A., {Christiansen}, J.~L., {Ciardi}, D.~R., {Cochran},
  W.~D., {Endl}, M., {Fortney}, J.~J., {Gautier}, III, T.~N., {Gilliland},
  R.~L., {Haas}, M.~R., {Hall}, J.~R., {Holman}, M.~J., {Howard}, A.~W.,
  {Howell}, S.~B., {Isaacson}, H., {Jenkins}, J.~M., {Klaus}, T.~C., {Latham},
  D.~W., {Li}, J., {Marcy}, G.~W., {Mazeh}, T., {Quintana}, E.~V., {Robertson},
  P., {Shporer}, A., {Steffen}, J.~H., {Windmiller}, G., {Koch}, D.~G., \&
  {Borucki}, W.~J. 2012, \nat, 481, 475

\end{thebibliography}

\end{document}